# A novel approach for fast mining frequent itemsets use N-list structure based on MapReduce

ARKAN A. G. AL-HAMODI[1*], SONGFENG LU[2]

[1]Research scholar, School of computer science Huazhong University of Science and Technology Wuhan 430074, PRC

[2]Associate professor, School of computer science Huazhong University of Science and Technology Wuhan 430074, PRC

E-mail: arkan_almalky@yahoo.com , lusongfeng@hust.edu.cn

**Abstract** Frequent Pattern Mining is a one field of the most significant topics in data mining. In recent years, many algorithms have been proposed for mining frequent itemsets. A new algorithm has been presented for mining frequent itemsets based on N-list data structure called Prepost algorithm. The Prepost algorithm is enhanced by implementing compact PPC-tree with the general tree. Prepost algorithm can only find a frequent itemsets with required (pre-order and post-order) for each node. In this chapter, we improved prepost algorithm based on Hadoop platform (HPrepost), proposed using the Mapreduce programming model. The main goals of proposed method are efficient mining frequent itemsets requiring less running time and memory usage. We have conduct experiments for the proposed scheme to compare with another algorithms. With dense datasets, which have a large average length of transactions, HPrepost is more effective than frequent itemsets algorithms in terms of execution time and memory usage for all min-sup. Generally; our algorithm outperforms algorithms in terms of runtime and memory usage with small thresholds and large datasets.

## 1 Introduction

Frequent pattern mining is one of the most important and popular research areas in mining Association rules field and data mining [1,2]. It is becoming the hot topic for finding frequent itemsets mining. Most of the proposed algorithms for frequent itemsets can be clustered in to Apriori method and FP-growth method. Repeatedly, the Apriori method scans the database to find frequent itemsets with generates a large set of a candidate [3]. FP-growth method scans the database twice to mines frequent itemsets without generating candidates [4]. The FP-growth uses FP-tree data structure to store database and employs a divide-and-conquer strategy to find frequent itemsets, which is much more efficient than Apriori method. In the frequent itemsets, two kinds of data structure (Node-list and N-list) have been proposed by Deng and et al. [5,6], to reduce the mining time and memory usage with mining frequent itemsets. The two of data structures based on a prefix tree with encoded nodes. The Node-list and N-list based on PPC-tree, and both of them consuming of memory because they need to encoding nodes with pre-order and post-order. Based on N-list algorithm called Prepost. In this chapter, we present a new method HPrepost algorithm based on PPC-tree under Mapreduce framework with Hadoop platform to obtain more efficiently for mining frequent itemsets, reduce running time, and usage memory.

## 2 Related work

The previous proposed algorithms for mining frequent itemsets divided into three groups, Generate candidate, frequent pattern growth and Hybrid approach. In recent years, three kinds of structure have been proposed for finding a frequent itemsets efficiently. Node-list structure was proposed by Deng and et al. [5], based on PPC-tree (pre-order post-order Code tree). N-List structure was proposed by Deng and et al. [6], needs to encode a node of the PPC-tree with pre-order and post-order. Both of the two structures are based on a prefix tree called PPC-tree. Two novel data structures are memory consuming because need to encode a node with pre-order and post-order. N-list based on algorithm for mining called NAFCP was proposed by Tuong Le and Bay Vo [7]. An enhanced N-list and Subsume-based algorithm for mining Frequent Itemsets (NSFI) algorithm that uses a hash table to improve the process of creating the N-lists associated with 1-itemsets and an enhanced N-list intersection algorithm was presented by Bay Vo and et al. [8]. New algorithm more effective with reducing the memory usage and mining time. An improved version of the mining top-rank-k frequent pattern (NTK) presents by Huynh et al. [9]. A hybrid algorithm based on PrePost proposed by Vo et al. [10], An improved PrePost algorithm uses a hash table to enhance the process of creating the N-lists data structure. Mapreduce programming framework is very well known technique for processing such massive of data [11,12]. Liao et al. [13] presented a parallel algorithm adapted for mining big data based on Hadoop platform under Mapreduce (MRPrepost). The algorithm employs N-list data structure, which improves PrePost by way of adding a prefix pattern. An improved Prepost algorithm with hadoop platform proposed by Thakare et al. [14]. The algorithm based on N-list data structure and enhanced by implementing compact PPC tree. Many researchers implement for finding frequent itemsets based on Mapreduce programming model [15-18].

## 3 Basic approaches

This section introduces relevant concepts and properties about PPC-tree, N-list and Prepost algorithm.

### 3.1 PPC-tree definition

Pre-order Post-order coding tree (PPC-tree) is a tree structure with satisfying where includes one root and set of nodes. Algorithm1 shows the pseudo-code of PPC-tree, and defined as following:

1) It consists of one root labeled “null," and a set of a node prefix sub-tree as the children of the root.
2) In the item prefix sub-tree, each node composed of five values: item-name, count, children-list, pre-order and post order. Item-name registers which item this node represents. Count registers the number of transactions presented by the portion of the path reaching this node. Children-list registers all children of the node. Pre-order is the pre-order rank of the node. Post-order is the post-order rank of the node.

PPC-tree looks like an FP-tree [19]. There are three important differences between them. First of all, PPC-tree is a simpler prefix tree because the FP-tree has a node-link field in each node and a header table structure to maintain the connection of nodes whose item-names are equal in the tree, where PPC-tree does not have such structures. Secondly, each node in the PPC-tree has pre-order and post-order fields while nodes in the FP-tree have not. The pre-order of a node is determined by a pre-order traversal of the tree. In a pre-order traversal, a node N is visited and assigned the pre-order rank before all its children are traversed recursively from left to right. In other words, the pre-order records the time when node N is accessed during the pre-order traversal. In the same way, the post-order of a node is determined by a post-order traversal of the tree. In a post-order traversal, a node N is visited and assigned its post-order rank after all its children have been traversed recursively from left to right. Finally, after an FP-tree is built, it will be used for frequent itemset mining during the total process of FP-growth algorithm, which is a recursive and complex process.

However, PPC-tree is only used for generating the Pre-Post code of each node. Later, we will find that after collecting the Pre-Post code of each frequent item first, the PPC-tree finishes its entire task and could be deleted.

**Example 1**, let the transaction database (TDB) shown by table1, will be used threshold (min-sup=0.3), the PPC-tree scan the datasets to determine F1= {a, b, c, d, e}, and removing items whose frequency does not satisfy {f, g}. The algorithm sorts F1 in descending order of support shown in table 2. Fig.1 show the PPC-tree constructs the root labeled “null," the number pair on the outside of the node is on the left pre-order and on the right post-order. The item node of (b) has counted is 5, with (1, 5) means that it is Pre-order is 1, and Post-order is 5.

### 3.2 N-list structure

Deng et al [6] presents a data structure named N-list for mining frequent itemsets. N-list generated from the prefix tree called PPC-tree, store nodes related to 1-items and are used to calculate the support of an item for FIM [20,21]. The N-list structure of a frequent itemset mining is a sequence of all the PP-codes registering the item node from the PPC-tree, with such a PP-codes arranged in the ascending order of their pre-order values. N-list is more compact than previous vertical structures.

**Definition 1 (PP code):** the PP code for each node N in a PPC-tree is ({N.pre-order, N.post-order}:N.count).

Fig. 2 show the N-list of all frequent itemsets, for example: N-list of item (d) includes two pp-code ({5, 2}:1) and ({8, 7}:2}.

**Table 1** Transaction datasets

| ID | Items |
|---|---|
| 1 | a, b, g |
| 2 | b, c, d, f, g |
| 3 | a, b, e |
| 4 | a, d |
| 5 | b, c, e |
| 6 | a, d, e, f |
| 7 | b, c |

**Table 2** Sort frequent items after removing infrequent 1-itemsets

| ID | items |
|---|---|
| 1 | b, a |
| 2 | b, c, d |
| 3 | b, a, e |
| 4 | a, d |
| 5 | b, c, e |
| 6 | a, d, e |
| 7 | b, c |

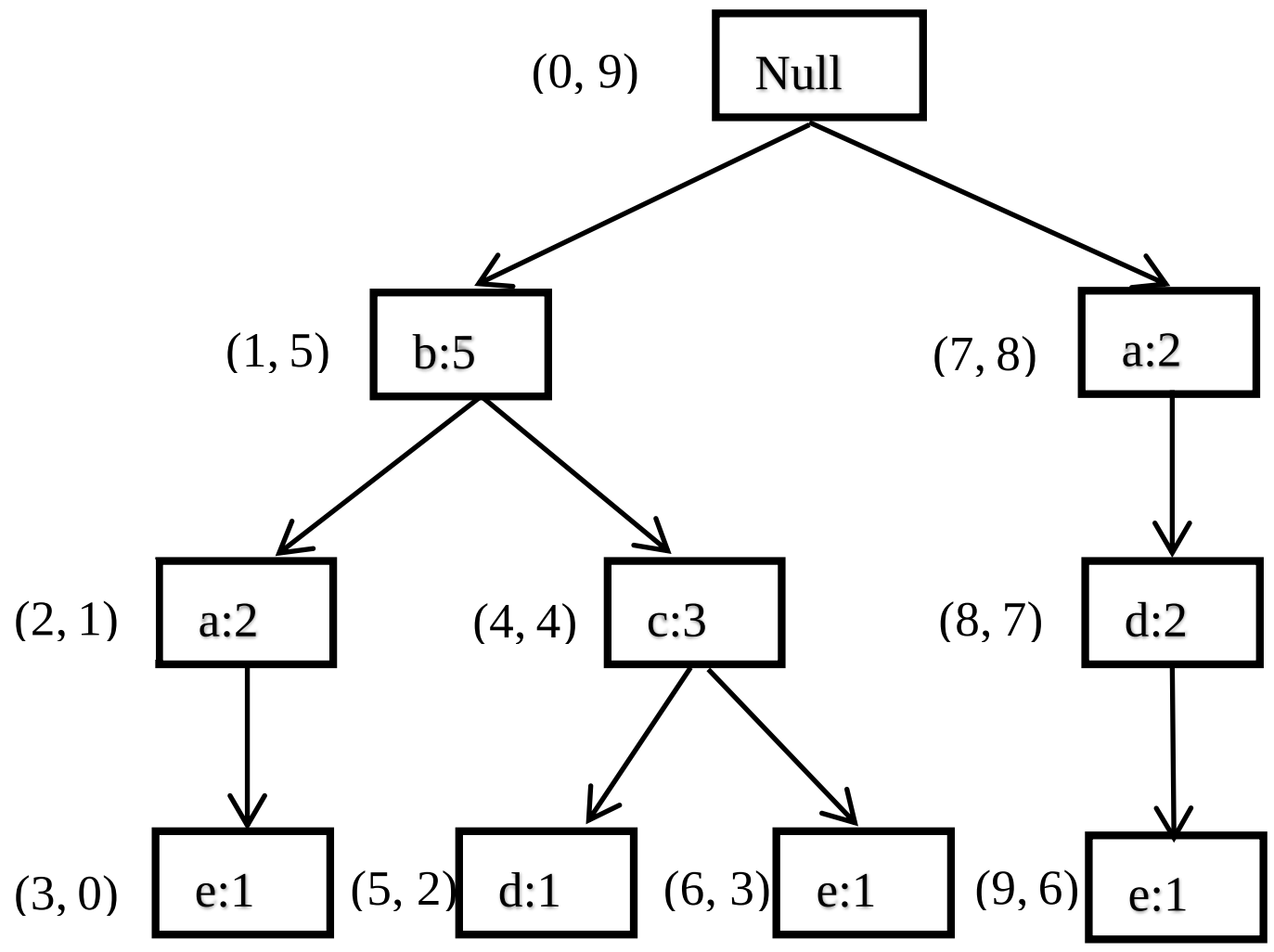

Fig.1 PPC-tree Construction

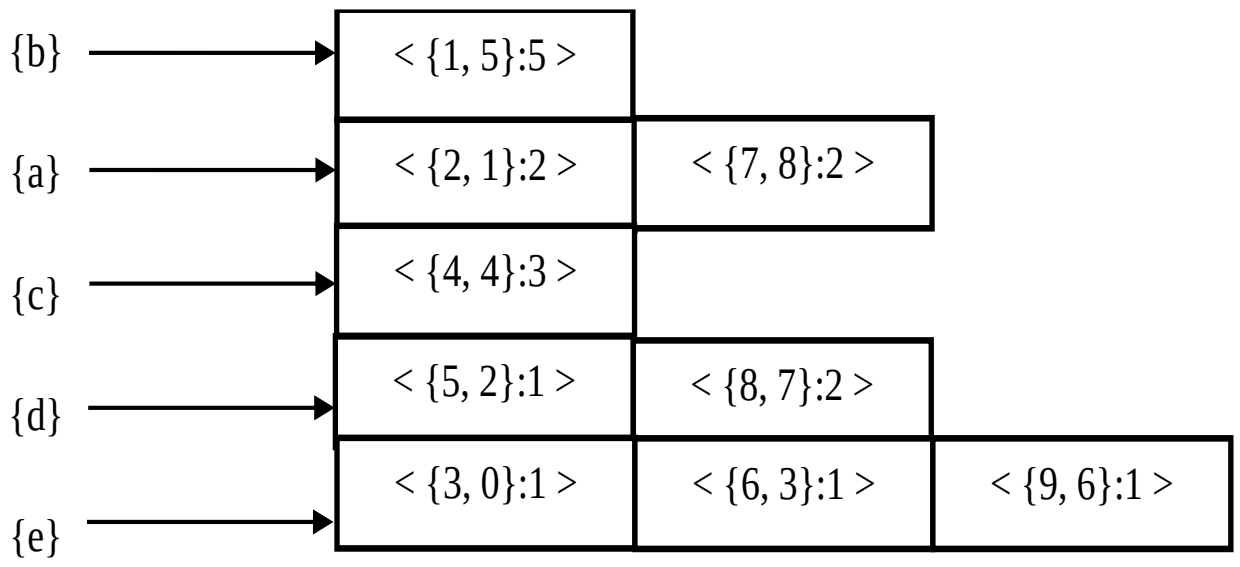

Fig.2 N-list of all frequent itemsets

### 3.3 Prepost data structure algorithm

Deng et al. [6] proposed a new algorithm called Prepost based on N-list data structure for mine frequent itemsets. Prepost algorithm finds frequent itemsets without generating candidate, and reduce count of support by an efficient strategy. It needs to scan a database twice to construct PPC-tree. The PPC-tree consists of one root labeled "Null," and asset of items prefix sub-tree as children. Each node consists of (item-name, count, children, Pre-order and Post-order).

The main steps of Prepost algorithm are:

1) Construct PPC-tree and identify all frequent 1-itemsets, each node in PPC-tree has a structure like (Pre-order, Post-order: count).
2) Based on PPC-tree, construct the N-list of each frequent 1-itemsets. By traversing The PPC-tree with Pre-order, we can access each node in PPC-tree. For each Node N, we insert < (N.Pre-order, N.Post-order): N.count> into the N-list of an item registered by N.
3) Scan PPC-tree to find all frequent 2-itemsets.
4) Mine all frequent K (>2)-itemsets.

Prepost algorithm uses N-list data structure for frequent itemsets mining. Prepost needs to scan the database twice to construct a PPC-tree and generate the N-list. The PrePost algorithm Scan DB and generate the FIM1, and in descending order according to the number of its support value to generate Fl. The PrePost algorithm Scan DB again, Select the frequent items in each record and arrange them in the order of F1, assuming the list of itemsets each record is [p|P], p is the first item on the list, P is the rest of the items. Call the function insert_tree ([p|P], Ti). Tree formed on the PP-tree, respectively pre-order traversal and post-order traversal, set pre-order and post-order of each node and establish N-list of 1-frequent itemsets. Mining frequent itemsets based N-list structure.

**Example 2**, for frequent itemsets (be), by the b, e merger. N-list of b is < (1, 5):5 >, and N-list of e is <(3, 0):1> <(6, 3):1> <(9, 6):1>, we find 1<3 and 5>0, this case (b) and (e) are on the same path, so the PP-Code: <(1, 5):1> added to the N-list of (be). Then traverse the next PP-Code of e <(6, 3):1>, 1 <6 and 5>3, the N-list <(1, 5):1> added to the N-list of (be), in which case the item sets (be) the N-list: <(1, 5):1> <(1, 5):1>, for the same pre-order, post-order to get the merger of PP-Code of be N-list structure <(1, 5):2>. Because 1<9, 5<6, this case, (e and b) are not on the same path, then traverse the next PP-Code of e.

## 4 HPrepost: The proposed method

The Prepost algorithm is a data mining algorithm for finding frequent itemsets which uses N-list data structure. We will use PPC-tree to improve the Prepost algorithm. We develop a high-performance Prepost structure approach over MapReduce programming model under Hadoop platform called (HPrepost), to process large amount of data and generate frequent itemsets. We use thee Map and Reduce function to find N-list of frequent 1-itemsets.

The following are the processing steps of HPrepost algorithm:

1) Dividing whole input data file into equal size block called sub-datasets, this process can be used in Hadoop. The Map function counts the data-Node in each block, and then generate frequent 1-itemset (F-list) according to the frequent support threshold.
2) Based on frequent 1-itemset, generate The F1-list, for each transaction of sub-datasets. And then sort frequent itemsets based on F1-list.
3) Generating a compressed tree called PPC-tree similar like FP-tree, and use method to find pre-order and post-order.
4) The pre-order and post-order rank of each node generate N-list of frequent 1-itemsets for each node in the tree.
5) Each Map function traversal every frequent itemsets in F1-list, until all frequent itemsets with the current prefix sub-tree. Based on prefix sub-tree generate frequent 2-itemsts.
6) Merge the prefix sub-tree with <p.pre-order, p.post-order, q.count> where P.pre-order <q.Pre-order and P.Post-order> q.post-order. Finally, reducing the grouped output.

Algorithm 1, shows pseudo-code for HPrepost algorithm. First of all, count the support values of each item in sub-itemsets. The counting of all items will be under map and reduce phases. Secondly, the mapper procedure generates frequent 1-itemsets using Prepost algorithm and then reduce procedure generates the PPC-tree, and then generate N-list of 1-itemsets. Each mapper is fed an HDFS block and input the Key (K) and Value (V). For each sub-itemsets, the mapper obtain on the sort of the values according to the F-list. The entered inputs (Key, value (index F-list) and value (sub-itemsets)) to reduce the procedure will build the PPC-tree.

**Algorithm 1:** HPrepost Algorithm

Input: A transection database (Ti) and min-sup (m).
Output: The set of frequent itemsets (F-list).
**Procedure** Mapper (K, V=Ti)
For each item $t_i$ in $T_i$ do
Output( <K=$t_i$, V=1>);
**End** Procedure
**Procedure** Reduce (K=$t_i$, V=S($t_i$))
Sum=0;
For each 1 in S($t_i$) do
Sum= sum +1;
If (sum>= m) then
output (<K=$t_i$, V=sum>)
else
Output F-list
**End** Procedure
**Procedure** Mapper (K, $T_i$)
For each item $t_i$ in $T_i$ do
Sort $t_i$ in $T_i$ according to the ordered of F-list
output a path [p|P] as the value <K, [p|P]>
**End** Procedure
**Procedure** Reduce (K, [p|P])
Create the root (null) of a PPC-tree
For each [p|P]
Call insert_tree([p|P], $T_i$)
Output PPC-tree
**End** Procedure

```
Function insert_tree([p|P], Ti)
If Ti has a child N (N.item-name= P.item-name) then
    N=N+1
Else
    Create a new node N, and add it to T children-list
    N=N+1
    If P is nonempty then
        Call inserte_tree (P, N)
End function
Procedure N-list (PPC-tree)
    For each node N of PPC-tree do
        Generate Pre Post order of each N
        Insert (N.Pre-order, N.Post-order): N.count) into N-list
End Procedure
```

## 5 Experiment results

In this section, we used four real datasets, which were often used in the previous study of mining frequent itemsets. These datasets are Chess, Mushroom, Pumsb and Kosarak. We downloaded them from http://fimi.ua.ac.be/data/. The characteristics of these databases are shown in Table 3, where show the number of items and transactions length. We have compared our proposed algorithm (HPrepost) with the original Prepost algorithm and FP-growth. The HPrepost, Prepost and FP-growth algorithms were all implemented in Java programming language. Each node runs on the Windows7 operating system. With JDK 1.8.0, java eclipse 4.4.2, Hadoop eclipse.

**Table 3** Characteristics of experimental datasets.

| Datasets | Items | Transaction | Avg. Length |
|---|---|---|---|
| Chess | 75 | 3196 | 37 |
| Mushroom | 119 | 8124 | 23 |
| Pumsb | 7117 | 49.046 | 74 |
| Kosarak | 41.270 | 990.002 | 8.1 |

### 5.1 Comparison of running time

The experimental results with respect to the running time of the compared algorithms HPrepost, Prepost and FP-growth with respect to the datasets have been shown in Table 3, chess, mushroom, Pumsb and Kosarak are presented in Fig. 3, 4, 5 and 6. In these Figures, the X axis is referred to the min-sup value, and the Y axis is referred to the total execution time. We analyzed the characteristics of them through the experimental results. As shown on the figures, the proposed algorithm guarantees the best execution time efficiency in almost all cases. Meanwhile, the FP-growth algorithm has the worst running time performance in almost all cases. HPrepost algorithm constructs PPC-tree structure, to extract all of the possible frequent itemsets. The Prepost algorithm spends a lot of time building the PPC-tree, and thus with a large min-sup, it is faster than FP-growth algorithm. However, with a small min-sup, HPrepost is much faster than Prepost and FP-growth in most cases. In Fig. 3, for the Chess dataset, the execution times of HPrepost are faster than Prepost and FP-growth algorithms increase rapidly when the threshold is decreased from 30% to 10%. FP-growth even cannot execute with a threshold < 10%. In Fig. 4, for the mushroom dataset, all the algorithms have a similar running time efficiency until the min-sup threshold > 25%. The execution time of FP-growth algorithm is consistently lower than those of HPrepost and Prepost algorithms. The proposed algorithm is more efficient than Prepost algorithm. In Fig. 5, the runtimes of HPrepost for Pumsb datasets are better than those of Prepost and FP-growth in most cases with all min-sup (10-50%). Prepost algorithm shows running time performance as well as that of FP-growth algorithm. In Fig. 6, the runtimes of HPrepost for Kosarak datasets are better than those of Prepost and FP-growth in most cases, especially with small thresholds. Prepost algorithm shows running time performance as well as that of FP-growth algorithm until the min-sup threshold <1%. In this regard, the proposed algorithm is more efficient than the others.

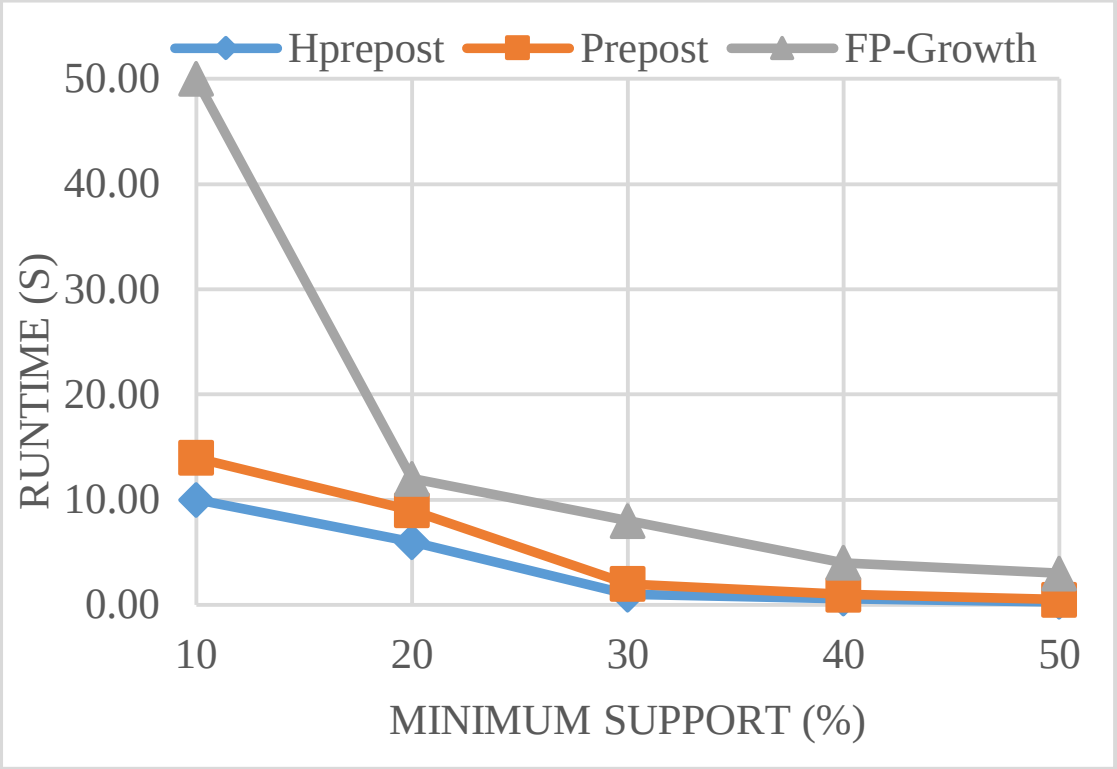

Fig. 3 Running time on the Chess datasets

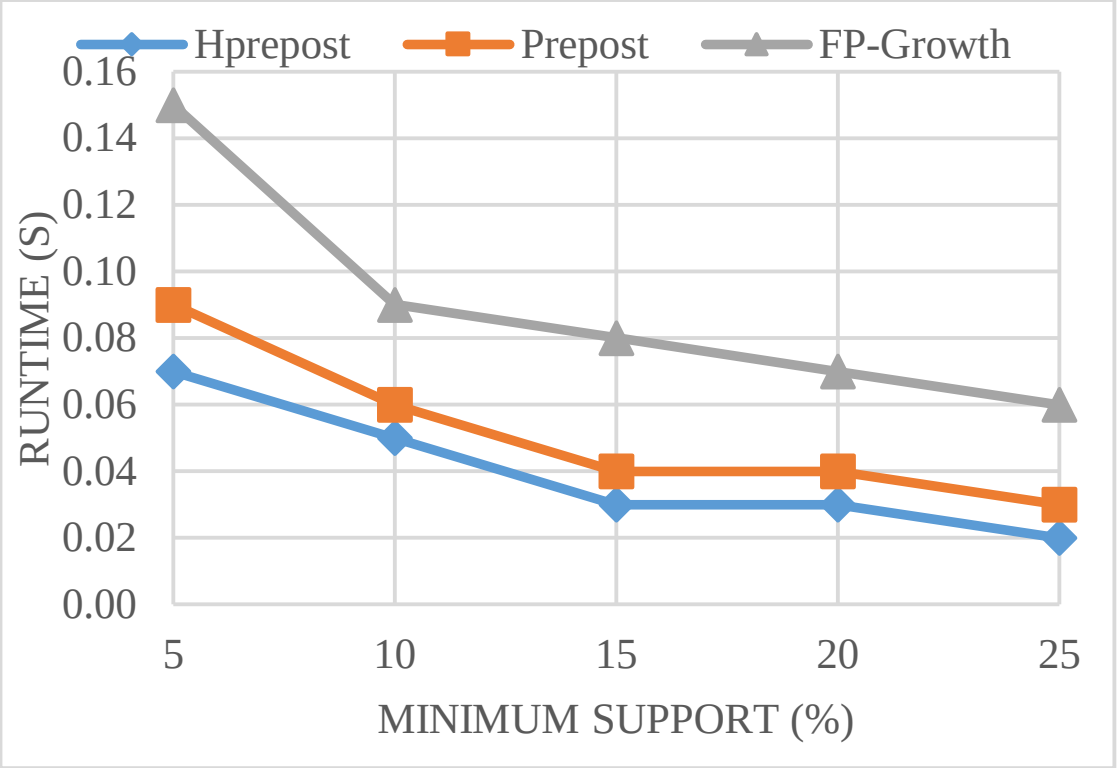

Fig. 4 Running time on the mushroom datasets

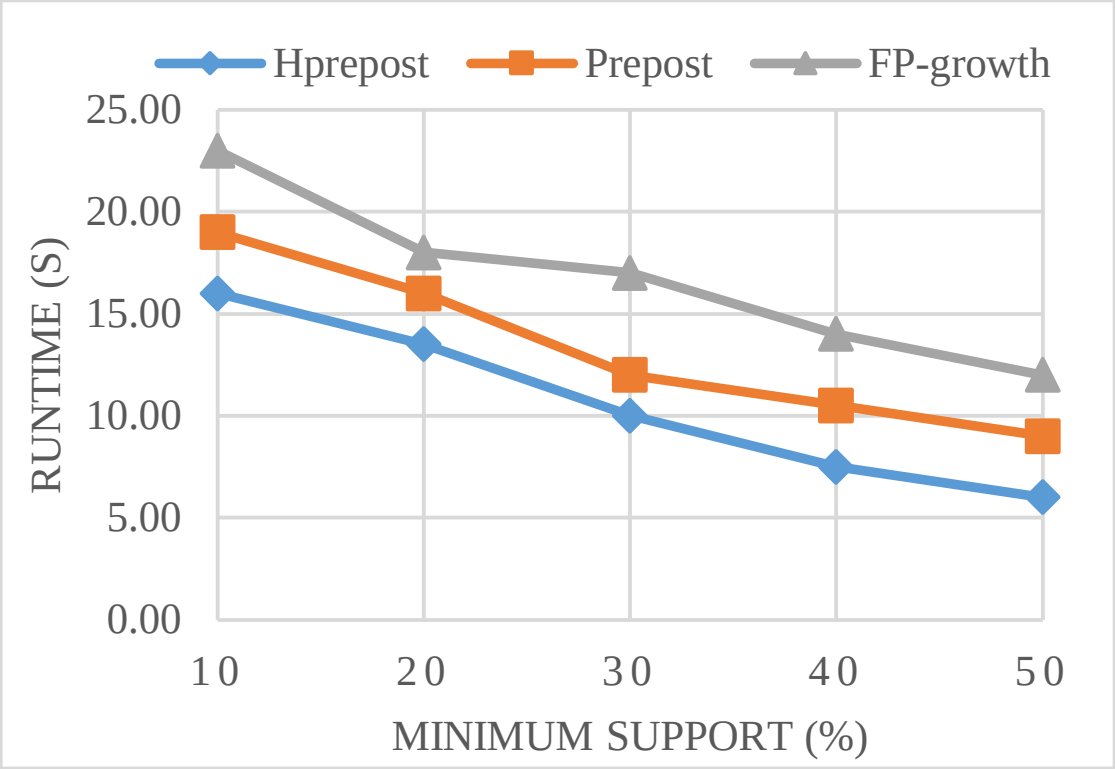

Fig. 5 Running time on the Pumsb datasets

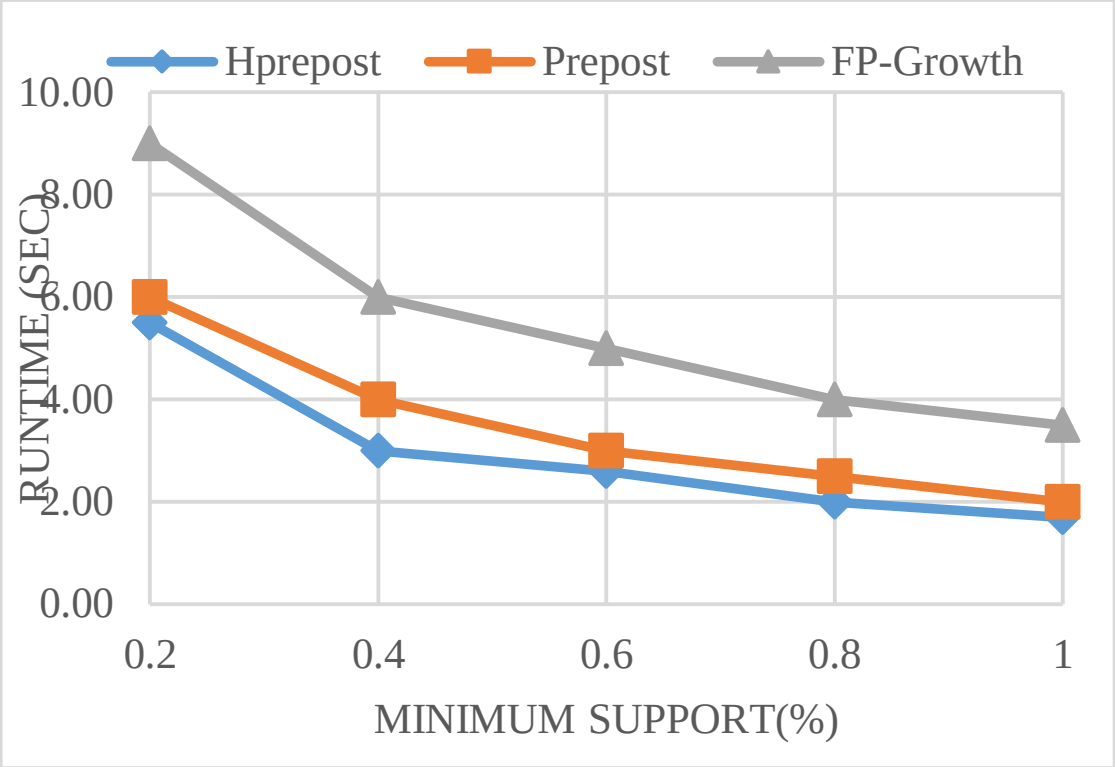

Fig. 6 Running time on the kosarok datasets

### 5.2 Memory consumption

In this section, memory usage tests are performed for the datasets in Table 3 as in the case of running time experiments, where the parameter settings for the tests are the same as those of the previous running time experiments. The experimental results with respect to the memory usage are presented in Fig. 7, 8, 9 and 10. Fig. 7 shows the memory required to store the user input data of proposed HPrepost, Prepost and FP-growth algorithms on the chess datasets. Chess is a very small dataset; necessary memory usage is also not large in the cases of Prepost and FP-growth. However, Prepost algorithm still requires large memory space to mine frequent itemsets when the threshold is less. The HPrepost algorithm uses the lowest amount of memory for all min-sup (10%, 20%, 30%, 40%, 50%). The HPrepost and FP-growth algorithms use less memory than Prepost algorithm when the min-sup is less.

The result of memory consumption for HPrepost, Prepost and FP-growth of mushroom datasets shown in Fig. 8. The Prepost algorithm performance becomes even worse as the threshold becomes lower than 15%. The HPrepost algorithm guarantees the most efficient memory usage for the given dataset and threshold settings. In Fig. 9 shows the results of memory used for HPrepost, Prepost and FP-growth algorithms on the Pumsb datasets. HPrepost algorithm also shows the most efficient and stable memory performance among them. Prepost suffers from memory overflow with respect to all of the threshold value. In Fig. 10 shows the results of memory used for HPrepost, Prepost and FP-growth algorithms on the Kosarak datasets. HPrepost algorithm also shows the most efficient and stable memory performance among them. Prepost suffers from memory overflow with respect to all of the threshold value. FP-growth algorithm uses the different tree structure called FP-tree. Prepost algorithm uses the N-list structure; the number of transactions is often greater than the number of nodes in the PPC-tree. Therefore, HPrepost algorithm generally requires less memory than do Prepost and FP-growth algorithms. In general, these experiments show that HPrepost is the best algorithm for mining frequent itemsets in terms of memory usage for all min-sup.

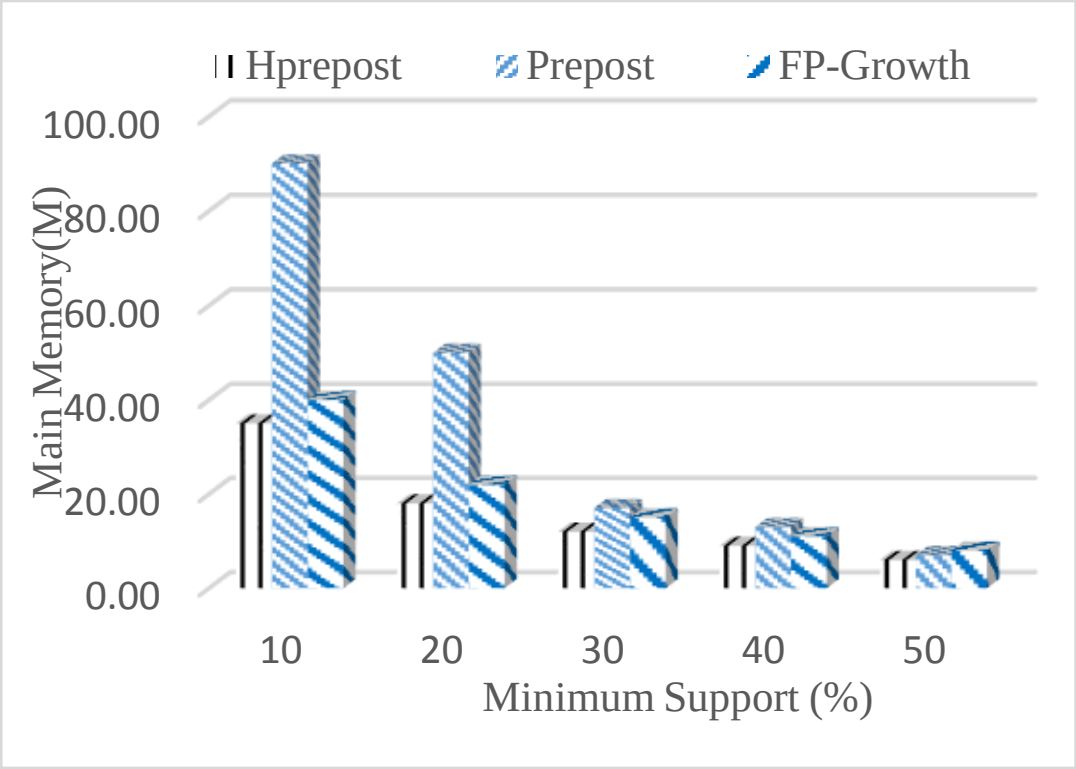

Fig. 7 Memory consumption on the Chess datasets

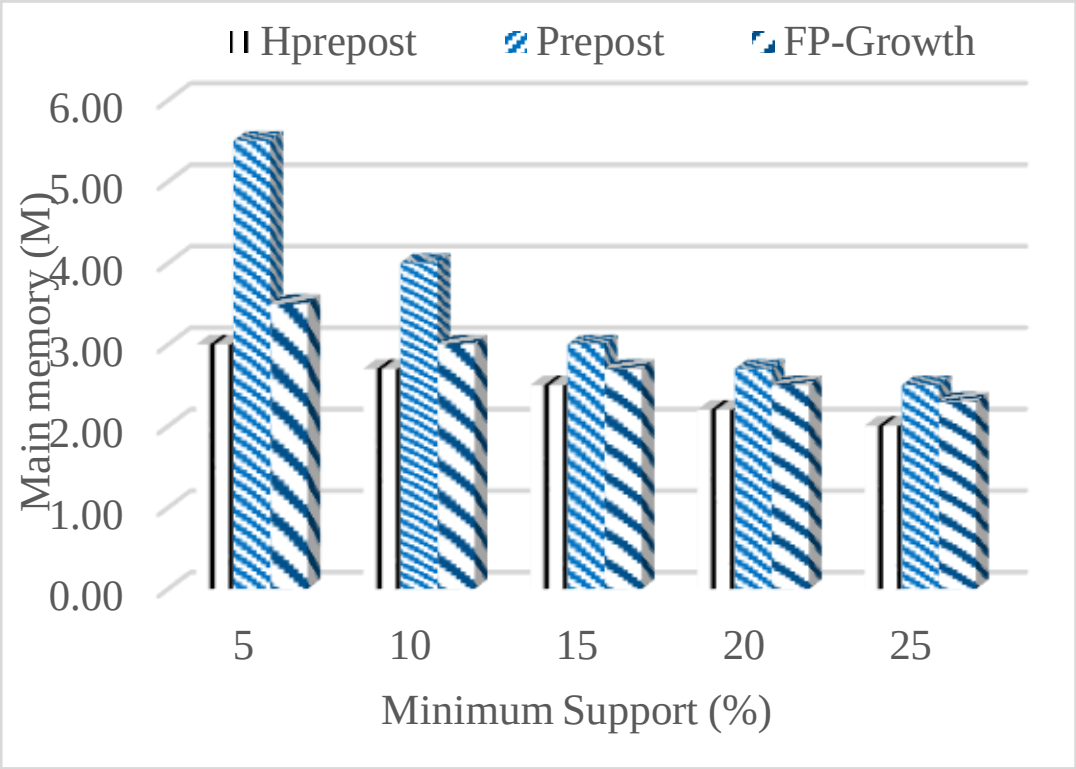

Fig. 8 Memory consumption on the Mushroom datasets

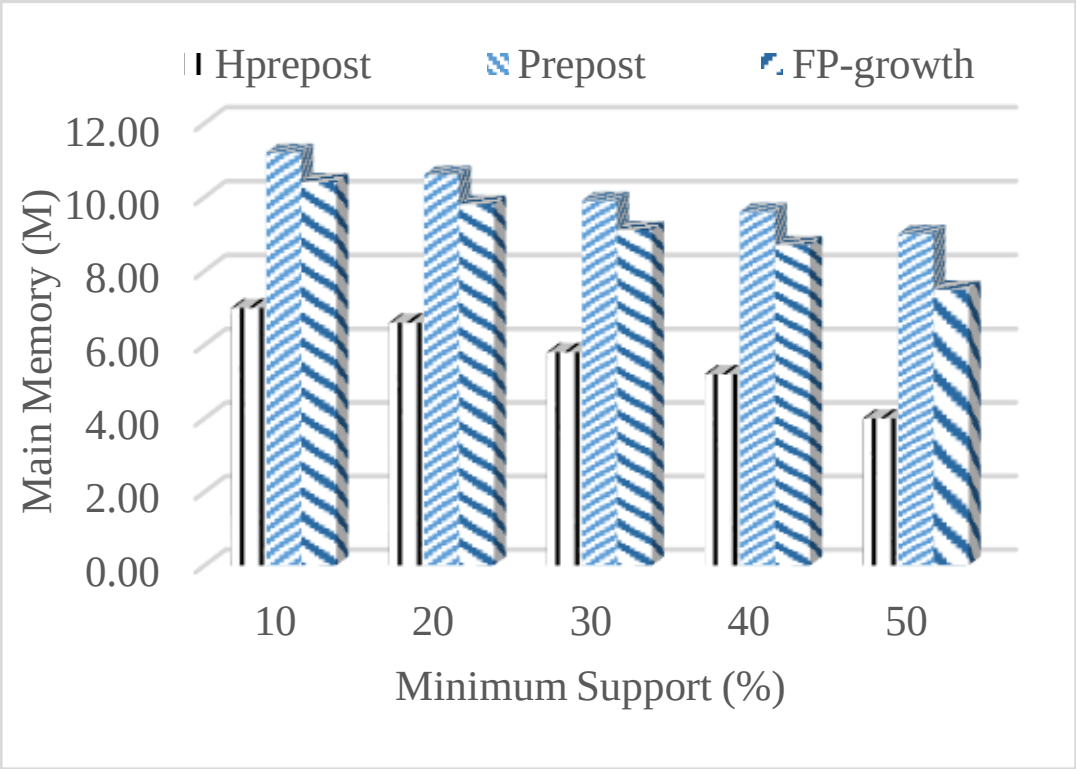

Fig. 9 Memory consumption on the pumsb datasets

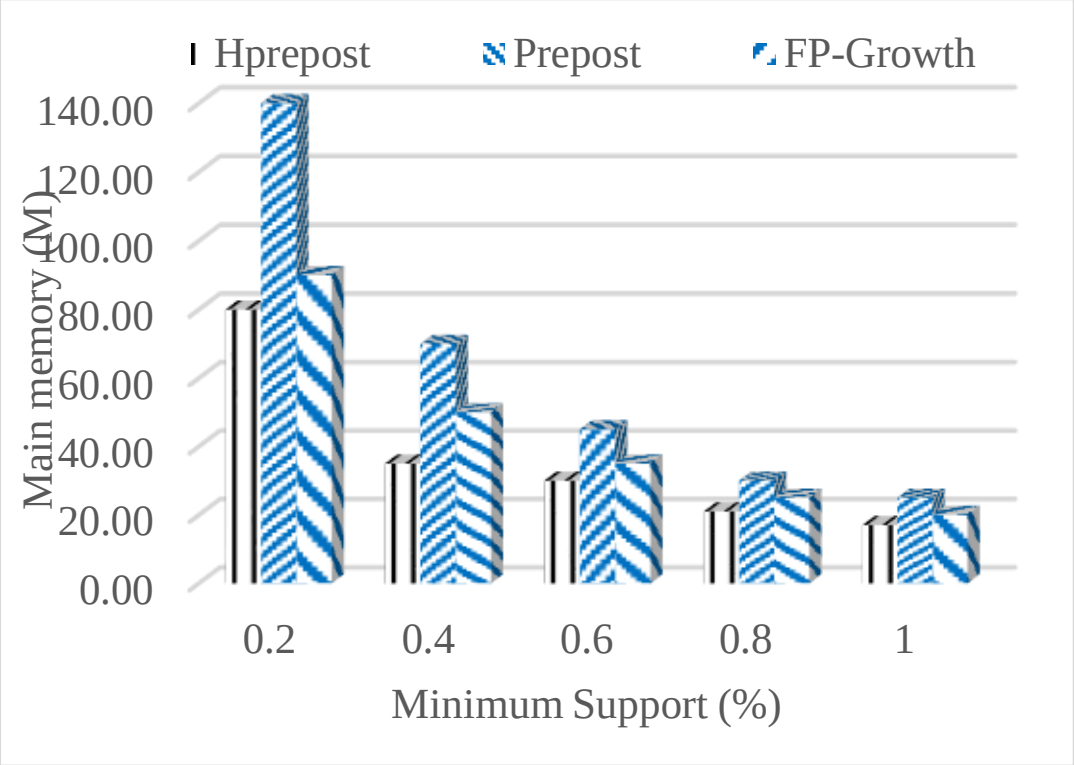

Fig. 10 Memory consumption on the Kosarok datasets

### 5.3 Discussion

As shown in the above sections, the main goals of this paper have been achieved by the proposed method algorithm. Results of Runtime and Memory usage shows all the algorithms have similar running time efficiency until the min-sup threshold is big. However, various empirical results of performance evaluation as the threshold becomes low, results proved that improvement of HPrepost algorithm becomes much efficient than Prepost and FP-growth algorithms. One of the main goals, has been achieved. Meanwhile, the second goal in this paper is to mine frequent itemsets mining results with large amount of data with all min-sup.

## 6 Conclusions

In this paper, we described a new algorithm in mining frequent itemsets based on N-list called HPrepost algorithm. We implemented the Prepost algorithm with Hadoop based on MapReduce. First, the paper describes step by step the Prepost algorithm. Second, Pre-Post Coding tree (PPC-tree). Finally, we described the N-list with four databases. The improved algorithm is faster than Prepost and FP-growth algorithms with all min-sup. Our algorithm is partition the transaction database, each partition can be executed to discover the frequent itemsets. Besides its novelty, using MapReduce makes this algorithm very easy to implement for finding frequent itemsets with large amount of datasets. The proposed algorithm shown that the usage memory consumption less than Prepost and FP-growth algorithm.

**Acknowledgments**

The authors gratefully acknowledge the support from the National Natural Science Foundation of China under (Grant No. 61173050).

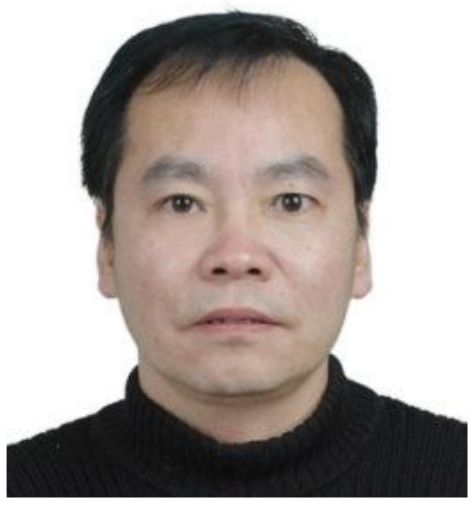

**SONGFENG LU** is working as an associate professor in School of Computer Science and Technology, Huazhong University of Science and Technology, China. He received Phd in Computer Science from Huazhong University of Science and Technology in 2001. His research areas include quantum computing, information security and data mining. You may contact him at lusongfeng@hust.edu.cn.

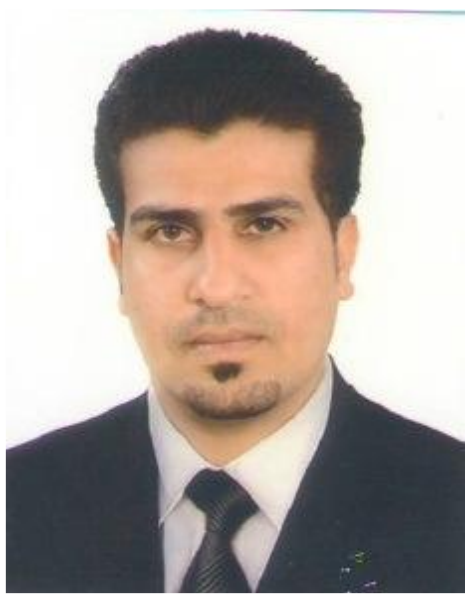

**ARKAN A. G. AL-HAMODI** is a Ph.D student in school of computer science, Huazhong University of Science & Technology, Wuhan, Hubei, P.R. china. He has completed M.Sc from the Department of Computer Science in S.H.I.A.T.S University, India in 2013. His research area of interest includes data mining and information technology. You may contact him at arkan_almalky@yahoo.com